\documentclass[pra,twocolumn,amsmath,amssymb,floatfix,reprint,footinbib,superscriptaddress,longbibliography,showkeys]{revtex4-1}
\usepackage{dcolumn}
\usepackage{graphicx}
\usepackage{mathrsfs}
\usepackage{mdwlist}
\usepackage{subfigure}
\usepackage{booktabs}
\usepackage{multirow}
\usepackage{amsmath}
\usepackage{textcomp}
\usepackage{upgreek}
\usepackage{dsfont}
\usepackage{amstext}
\usepackage{amssymb}
\usepackage{amsbsy}
\usepackage{appendix}
\usepackage{soul}
\usepackage{threeparttable}
\usepackage{bbm}
\usepackage{svg}
\usepackage{amsthm}
\usepackage{graphicx}
\usepackage{lipsum}
\usepackage{xcolor}
\usepackage[none]{hyphenat}
\usepackage{textcomp}
\usepackage{color}
\usepackage[colorlinks,citecolor=blue]{hyperref}
\usepackage{subfigure}
\definecolor{Dgreen}{RGB}{0, 100, 0}
\usepackage{url}
\usepackage[colorlinks]{hyperref}
\usepackage{array,xcolor}
\hypersetup{%
	plainpages=true,
	breaklinks=true,  
	hypertexnames=false,  
	pageanchor=true,
	colorlinks=true,
	linkcolor={blue},
	citecolor={red},
	urlcolor={blue},
	anchorcolor={black}
}

\begin{document}
\title{Engineering dissipation and control pulses for high-fidelity fault-tolerance quantum computing}
\author{Shao-Wei Xu}
\affiliation{Fujian Key Laboratory of Quantum Information and Quantum Optics, Fuzhou University, Fuzhou 350116, China}
\affiliation{Department of Physics, Fuzhou University, Fuzhou 350116, China}
\author{Zhe-Yuan Zhang}
\affiliation{Department of Physics, Fuzhou University, Fuzhou 350116, China}
\author{Yi-Tong Shi}
\affiliation{Department of Physics, Fuzhou University, Fuzhou 350116, China}
\author{Ye-Hong Chen}\thanks{yehong.chen@fzu.edu.cn}
\affiliation{Fujian Key Laboratory of Quantum Information and Quantum Optics, Fuzhou University, Fuzhou 350116, China}
\affiliation{Department of Physics, Fuzhou University, Fuzhou 350116, China}
\affiliation{Quantum Information Physics Theory Research Team, Center for Quantum Computing, RIKEN, Wako-shi, Saitama 351-0198, Japan}
\affiliation{Institute of Quantum Science and Technology, Yanbian University, Yanji 133002, China}
\author{Yan Xia}\thanks{xia-208@163.com}
\affiliation{Fujian Key Laboratory of Quantum Information and Quantum Optics, Fuzhou University, Fuzhou 350116, China}
\affiliation{Department of Physics, Fuzhou University, Fuzhou 350116, China}
\affiliation{Institute of Quantum Science and Technology, Yanbian University, Yanji 133002, China}

\begin{abstract}
Cat-state qubits, a prominent class of bosonic encodings, offer a promising pathway toward hardware-efficient fault-tolerant quantum computing. 
In this manuscript, we propose an optimally robust control protocol for the cat-state qubits which are stabilized by engineering two-photon dissipation. By deriving 
an effective two-level description in the cat-state subspace and applying shortcut-to-adiabaticity via inverse engineering, we design a robust protocol to achieve fast and 
high-fidelity state transfer in the cat-state qubit. We analyze the sensitivity to systematic control errors and identify an optimal robustness condition that strongly 
suppresses errors induced by imperfections in the driving fields. Furthermore, we show that dissipative confinement efficiently suppresses leakage out of the cat-state 
subspace caused by the pure dephasing, highlighting an intrinsic advantage of dissipative-cat qubits. This work establishes a robust and leakage-suppressing framework for 
high-fidelity bosonic qubit control, offering a promising route toward scalable fault-tolerant quantum computing. 
\end{abstract}

\date{\today}
\maketitle

\section{Introduction}
Quantum computing holds the promise of surpassing classical computers on certain problems, but its physical realization is inherently fragile and prone to decoherence and errors 
caused by environmental noise \cite{Hidary2019,Kockum2019,Lipton2021,Shor1994,Grover1996,Zurek2003,Braun2001PRL,Carvalho2004PRL,HornbergerPRA}. 
To overcome this challenge and achieve reliable, large-scale quantum information processing, quantum error correction (QEC) has emerged as an indispensable cornerstone \cite{Shor1995PRA,Steane1996PRL,Terhal2015}. 
By systematically encoding logical information across multiple physical systems, QEC protocols can detect and correct errors without disturbing the stored quantum state. Among 
various QEC approaches, bosonic encodings provide an efficient and promising pathway by encoding information within the large Hilbert space of a single bosonic mode \cite{Ma2021,Gottesman2001PRA,Mirrahimi2014NJP,Michael2016PRX,Cai2021,Joshi2021,Yan2026PRL}. 
In this approach, error correction is achieved by controlling excitation manifolds rather than increasing the number of physical qubits \cite{Google2023,Fowler2012PRA}. 
As a result, bosonic codes can significantly reduce hardware overhead \cite{Cai2021,Joshi2021,Ofek2016}. 

A prominent implementation of bosonic encoding is the cat-state qubits, where logical information is encoded in superpositions of photonic coherent states \cite{Cochrane1999PRA,Mirrahimi2014NJP,Vlastakis2013,Zheng2023PRL,Li2024PRA,Yan2026PRA,Xu2026CP,Xiao2026PRA}. 
Cat-state qubits possess an intrinsic noise bias. As the coherent amplitude increases, phase-flip errors are exponentially suppressed, while bit-flip errors 
remain dominant \cite{Lescanne2020,Guillaud2019PRX,Chamberland2022PRXQu}. 
This strong bias enables simplified QEC strategies that primarily target bit-flip errors, reducing the number of required error correction layers and making 
cat-state qubits particularly attractive for fault-tolerant quantum computation \cite{Guillaud2019PRX,Chamberland2022PRXQu,Puri2020}. 

Two widely used mechanisms for stabilizing cat-state qubits are engineered Kerr nonlinearity \cite{Puri2017npjQI,Puri2019PRX,Chen2022PRAppl,Kang2022PRR,Grimm2020,Ding2025NC} 
and engineered two-photon dissipation \cite{Mirrahimi2014NJP,Leghtas2015,Reglade2024,Gautier2023PRXQu}, resulting in Kerr-cat qubits and dissipative-cat qubits, respectively. 
Both schemes have been successfully implemented in superconducting circuit platforms, and experimental progress has been made in the preparation, stabilization, and 
readout of cat-state qubits \cite{Grimm2020,Ding2025NC,Leghtas2015,Reglade2024}. The fault-tolerance advantages of cat-state qubits, however, are predicated on the ability to 
control them with high fidelity, which is a prerequisite that remains elusive due to several control challenges \cite{Ma2021,Cai2021}. 

In particular, control fields are inevitably subject to systematic imperfections and noise, which can significantly reduce gate fidelities. Moreover, control-induced leakage 
out of the cat-state subspace, along with the interplay between dissipation and coherent control, further complicate the design of reliable control protocols \cite{Puri2020,Chamberland2022PRXQu}. 
In this manuscript, we address these challenges by proposing an optimally robust control protocol for dissipative-cat qubits. 
By deriving an effective two-level description within the cat-state subspace, we design fast population transfer protocols based on 
shortcut-to-adiabaticity (STA) methods \cite{Berry2009,Guery2019,Chen2010PRL,Lewis1969,Chen2011PRA,Dridi2020PRL}. 
Specifically, we employ a STA protocol via inverse engineering of the state trajectory. 
First, an evolution path for the logical state is predefined on the Bloch sphere \cite{Chen2010PRL}. 
The corresponding control Hamiltonian is then obtained by mapping this trajectory onto the required time-dependent drive parameters \cite{Guery2019,Chen2010PRL,Chen2010PRL,Lewis1969,Chen2011PRA}. 
This strategy enables a fast and high-fidelity state transfer that circumvents the speed limitations and stringent requirements of conventional adiabatic evolution \cite{Guery2019,Berry2009}. 

We further analyze the sensitivity of the protocol to systematic control errors and identify an optimal robustness condition that strongly suppresses errors induced by 
imperfections in the single-photon drive. In addition, we show that two-photon dissipation confinement plays a crucial role in suppressing leakage induced by pure dephasing, 
highlighting an intrinsic advantage of dissipative cat qubits. Numerical simulation results confirm the robustness and feasibility of the proposed approach under realistic 
decoherence. Together, these results establish a comprehensive framework that integrates fast control, error resilience, and leakage suppression, offering a promising pathway 
toward reliable manipulation of bosonic qubits in fault-tolerant quantum computing architectures.

\section{Two photon dissipation confinement}
\begin{figure}
	\centering
	\scalebox{0.55}{\includegraphics{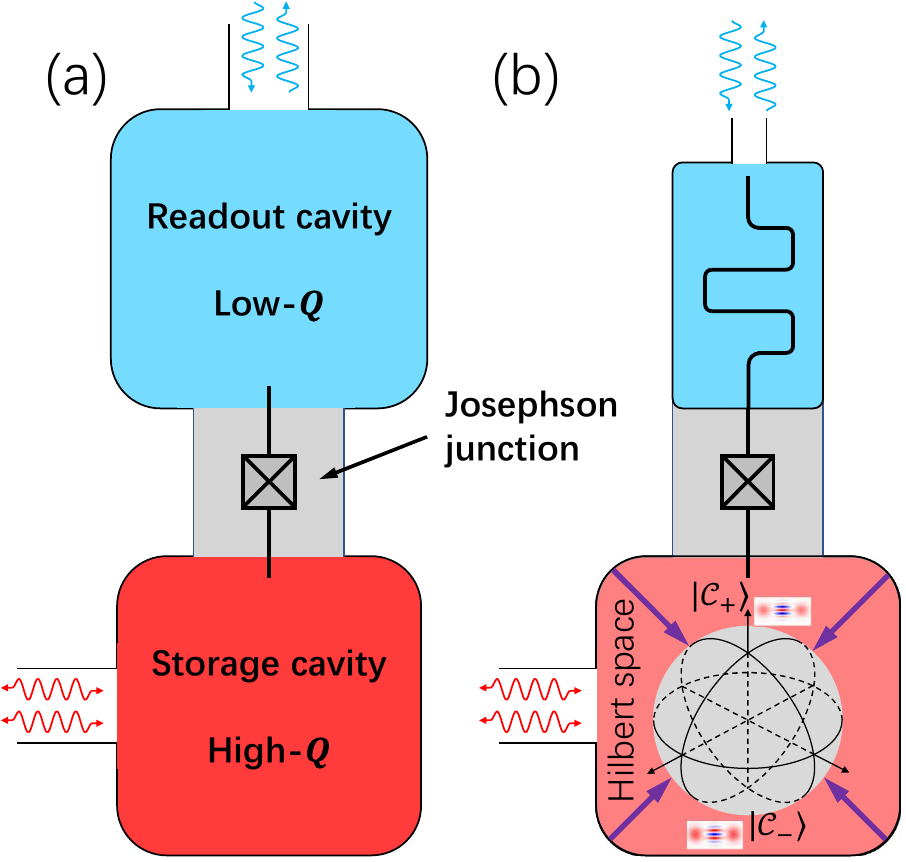}}
	\caption{(a) Schematic of the setup: Two cavities are coupled through a Josephson junction. A weak resonant drive $\varepsilon_b$ and 
				a strong off-resonant pump $\varepsilon_p$ are applied to the readout cavity, creating the appropriate nonlinear interaction, 
				which generates a coherent superposition of steady states in the storage cavity. 
	 		 (b) Confinement of a quantum state belonging to a large Hilbert space into cat-state subspace $\left\{|\mathcal{C}_{\pm}\rangle\right\}$. As indicated by the 
			 purple arrows, stabilizing forces direct all states toward the cat-state subspace.}	
	\label{Sch}
\end{figure}

Considering a model that two cavities, as shown in Fig.~\ref{Sch} (a), a storage mode (high-$Q$) and a readout mode (low-$Q$), are coupled by a Josephson junction. 
Under the rotating-wave approximation (RWA), the system Hamiltonian reads \cite{Mirrahimi2014NJP,Leghtas2015,Reglade2024} (hereafter $\hbar = 1$) 
\begin{align}
	H_{ab} = g^* a^2 b^\dagger + g a^{\dagger 2} b - \varepsilon_b^* b - \varepsilon_b b^\dagger, 
\end{align}
where $a$ ($a^\dagger$) and $b$ ($b^\dagger$) are the annihilation (creation) operators of storage and readout cavities, respectively. 
$g$ denotes the hopping rate between storage and readout cavities, and $\varepsilon_b$ is the amplitude of a resonant drive applied to the readout mode. 
For simplicity, we assume that $g$ and $\varepsilon_b$ are real. 

The system dynamics is well described by the following Lindblad master equation 
\begin{align}
	\dot{\rho} =& -i [H_{ab},\rho] + \kappa_a \mathcal{D}[a] \rho + \kappa_b \mathcal{D}[b] \rho \cr
					 & + \kappa_a^\phi \mathcal{D}[a^\dagger a] \rho + \kappa_b^\phi \mathcal{D}[b^\dagger b] \rho, 
\end{align}
where $\rho$ is the joint density matrix and $\mathcal{D}[o] \rho = o \rho o^\dagger - \frac{1}{2} \left\{ o^\dagger o, \rho \right\}$ is the standard Lindblad superoperator. 
$\kappa_a$ ($\kappa_a^\phi$) and $\kappa_b$ ($\kappa_b^\phi$) are the damping rates (pure dephasing rates) of storage and readout cavities, respectively. 

In the parameter regime $\left\{\kappa_a^\phi,\ \kappa_b^\phi\right\} \ll \kappa_b$, we temporarily disregard the influence of the pure dephasing. 
Assuming $\left\{g,\ \varepsilon_b,\ \kappa_a\right\} \ll \kappa_b$, the mean photon number in the readout mode is always much smaller than one \cite{Leghtas2015}. 
Therefore, we can obtain a master equation for the reduced density matrix of the storage mode by adiabatically eliminating the readout mode (see Appendix~\ref{Eliminate} for 
more details)
\begin{align}\label{Eq3}
	\dot{\rho}_{a} &\approx \frac{\kappa_2 \alpha^2}{2} [a^{\dagger 2}-a^2, \rho_{a}] + \kappa_2 \mathcal{D}[a^2] \rho_{a} + \kappa_a \mathcal{D}[a] \rho_{a} \cr
				   &= \kappa_2 \mathcal{D}[a^2 - \alpha^2] \rho_{a} + \kappa_a \mathcal{D}[a] \rho_{a}, 
\end{align}
where $\kappa_2=4g^2/\kappa_b$ is the effective two-photon dissipation rate, $\alpha = \sqrt{\varepsilon_b / g}$ is the coherent amplitude.

The steady-state solutions satisfy $\dot{\rho}_{a}=0$. In the absence of the damping in storage mode, the dissipative dynamics governed by Eq.~(\ref{Eq3}) naturally 
stabilize the system into a steady-state manifold. Specifically, the steady states are determined by the jump operator $(a^2 - \alpha^2)$. 
Since the coherent states $\lvert\pm\alpha\rangle$ satisfy $(a^2 - \alpha^2)\lvert\pm\alpha\rangle=0$, the Lindblad superoperator vanishes on these states. 
Consequently, any superposition of the coherent states $\lvert \pm \alpha \rangle$, specifically the even and odd cat-states $|\mathcal{C}_\pm\rangle$, 
forms a degenerate steady-state subspace protected by the engineered two-photon dissipation 
\begin{align}
	|\mathcal{C}_\pm\rangle = N_{\pm} (\lvert\alpha\rangle \pm \lvert-\alpha\rangle), 
\end{align}
where $N_{\pm} = 1 / \sqrt{2(1 \pm e^{-2|\alpha|^2})}$ are normalized coefficients and $\lvert\pm \alpha \rangle = D(\pm \alpha) |0\rangle$ is the coherent state. 
Here, $D(\pm \alpha)=\exp[\pm \alpha (a^\dagger-a)]$ is the displacement 
operator. The excited states can be represented as $|\psi_\pm^{e,n}\rangle = N_\pm^{e,n} [D(\alpha) \pm D(-\alpha)] |n\rangle$, where $|n\rangle$ are the Fock states.

{
The two-photon dissipation $\mathcal{L}_{\rm 2ph} \equiv \kappa_2 \mathcal{D}[a^2 - \alpha^2]$ defines a non-trivial Liouvillian spectrum. The logical subspace $\{|C_\pm\rangle\}$ resides in the 
steady-state manifold (zero eigenvalue). In the limit of a large $\alpha$, the dynamics near the steady-state manifold can be linearized by introducing the displacement 
transformation $a = \alpha + f$, where $f$ represents the small quantum fluctuation around the steady-state. Expanding the jump operator in $\mathcal{L}_{\rm 2ph}$, we have 
\begin{align}
	 a^2 - \alpha^2 =& (\alpha + f)^2 - \alpha^2 \cr
	 				=& 2\alpha f + f^2. 
\end{align}
For a large $\alpha$, we can safely neglect the higher-order term $f^2$, then obtain $ a^2 - \alpha^2 \approx 2\alpha f$.~The two-photon dissipation term then reduces 
to an effective single-photon loss process 
\begin{align}
	\kappa_2 \mathcal{D}[a^2 - \alpha^2]\rho \approx \kappa_{\rm eff} \mathcal{D}[f]\rho, 
\end{align}
where $\kappa_{\rm eff} = 4\kappa_2 |\alpha|^2$ is the effective decay rate. The Liouvillian gap \cite{Mori2023PRL, Minganti2018RPA}
\begin{align}
	\Delta_L = \frac{\kappa_{\rm eff}}{2} = 2\kappa_2 |\alpha|^2, 
\end{align}
which determines the rate at which the system is restored to the cat-state subspace after a perturbation, is given by the non-zero eigenvalue of the dissipative 
superoperator $\mathcal{L}_f \equiv \kappa_{\rm eff} \mathcal{D}[f]$ with the smallest magnitude. 
}

\section{Control protocol for cat-state qubit}
\subsection{Control Hamiltonian}\label{Hc}
The two cat states can be fliped by the action of annihilation operator $a$, i.e., 
\begin{align}
	a|\mathcal{C}_\pm\rangle &= \frac{N_\pm}{N_\mp} \alpha |\mathcal{C}_\mp\rangle. 
\end{align}
For large $\alpha$, we can apply a single-photon drive to the storage cavity to achieve the bit-flip of cat states, i.e., 
$a|\mathcal{C}_\pm\rangle \simeq \alpha |\mathcal{C}_\mp\rangle$. 

Applying a single-photon drive can achieve the population transfer of the cat-state qubit, the control Hamiltonian reads 
\begin{align}
	H_0(t) = \varepsilon_d(t) (a^\dagger + a), 
\end{align}
where $\varepsilon_d(t)$ is the single-photon drive amplitude. 
{
The corresponding fluctuation Hamiltonian can be approximated as $H_f(t) \simeq \varepsilon_d(t) (f^\dagger + f)$, which may induce leakage out of the cat-state subspace. 
Therefore, to ensure that the system remains stabilized within the cat-state subspace $\left\{|\mathcal{C}_{\pm}\rangle\right\}$, the drive amplitude should satisfy 
$\varepsilon_d(t) \ll \Delta_L = 2\kappa_2 |\alpha|^2$. 
}

Encoding quantum information into the cat states, we can then define $\sigma_+ = |\mathcal{C}_+\rangle \langle\mathcal{C}_-|$ and 
$\sigma_- = |\mathcal{C}_-\rangle \langle\mathcal{C}_+|$ to be the raising and lowering operator, respectively. Therefore, the Pauli operators can be represented as 
\begin{align}
	\sigma_x &= \sigma_+ + \sigma_-, \cr
	\sigma_y &= i (\sigma_- - \sigma_+), \cr
	\sigma_z &= \sigma_+\sigma_- - \sigma_-\sigma_+. 
\end{align}
We can further obtain the following expressions 
\begin{align}\label{Eq7}
	P_\mathcal{C} a P_\mathcal{C} &= \alpha \left[ \frac{A+A^{-1}}{2} \sigma_x + i \frac{A-A^{-1}}{2} \sigma_y \right],\cr
	P_\mathcal{C} a^\dagger a P_\mathcal{C} &= |\alpha|^2 \left[ \frac{A^2+A^{-2}}{2} \mathbbm{1} - \frac{A^2-A^{-2}}{2} \sigma_z \right], 
\end{align}
where $P_\mathcal{C} = |\mathcal{C}_+\rangle \langle\mathcal{C}_+| + |\mathcal{C}_-\rangle \langle\mathcal{C}_-|$ is the projection operator, $A = N_-/N_+$ is a 
dimensionless constant, and $\mathbbm{1}$ denotes the unit matrix in the cat-state subspace $\left\{|\mathcal{C}_{\pm}\rangle\right\}$. Projecting $H_c(t)$ onto the cat-state 
subspace, following Eq.~(\ref{Eq7}), the effective control Hamiltonian can be expressed as 
\begin{align}\label{H0}
	H_0^{\rm eff}(t) \simeq& \varepsilon_d(t) \alpha (A+A^{-1}) \sigma_x \cr
		   =& \frac{1}{2}
		   \begin{pmatrix}
			0& \Omega_R(t) \\[6pt]
			\Omega_R(t)& 0
		   \end{pmatrix}, 
\end{align}
where $\Omega_R(t) = 2(A+A^{-1}) \alpha \varepsilon_d(t)$ is the effective Rabi frequency. 

In general, the flipping of the two cat states can be achieved through Eq.~(\ref{H0}). {However, in actual experiments, this single-photon drive field may 
have parameter imperfections arising from amplitude and phase instabilities of the microwave pulses delivered to the storage cavity. These imperfections can originate 
from several experimental factors, including thermal drift in cryogenic coaxial lines, the finite resolution of arbitrary waveform generators, and local-oscillator 
leakage or gain fluctuations in I/Q mixers. Since these imperfections are typically quasi-static during the experimental timescale, they are generally regarded as 
systematic errors and therefore produce the same error each time the control pulse is applied \cite{Krantz2019APR}. 

Now, we denote the single-photon drive strength in 
the following form,} 
\begin{align}
	\varepsilon_d(\lambda,t) &= (1+\lambda) \varepsilon_d(t), 
\end{align}
where, $\lambda$ is an error rate. The corresponding Hamiltonian becomes 
\begin{align}
	H_\lambda(t) = (1+\lambda) H_0(t). 
\end{align}
In the cat-state subspace $\left\{|\mathcal{C}_{\pm}\rangle\right\}$, the erroneous Hamiltonian can be expressed as 
\begin{align}
	H_\lambda^{\rm eff} &= H_0^{\rm eff} + \lambda H', \cr
	H' &= \frac{1}{2}
		   \begin{pmatrix}
			0& \Omega_R \\[6pt]
			\Omega_R& 0
		   \end{pmatrix}, 
\end{align}
where $H_0^{\rm eff} \equiv H_0^{\rm eff}(t)$ and $\Omega_R \equiv \Omega_R(t)$.

\subsection{Error sensitivity analysis}\label{Analysis qs}
Assuming that, any state within the cat-state subspace $\left\{|\mathcal{C}_{\pm}\rangle\right\}$ can, without loss of generality, be expressed as 
\begin{align}\label{psi_t}
	|\psi(t)\rangle =&\ \cos\frac{\theta}{2} e^{i\phi / 2} e^{-i \zeta /2} |\mathcal{C}_+\rangle \cr
					& + \sin\frac{\theta}{2} e^{-i\phi / 2} e^{-i \zeta /2} |\mathcal{C}_-\rangle \cr
					=& \begin{pmatrix}
						\cos\frac{\theta}{2} e^{i\phi / 2} \\[6pt]
						\sin\frac{\theta}{2} e^{-i\phi / 2}
					   \end{pmatrix}
						e^{-i \zeta /2}, 
\end{align}
and the corresponding orthogonal state is given by
\begin{align}
	|\psi_\bot(t)\rangle = \begin{pmatrix}
						-\sin\frac{\theta}{2} e^{i\phi / 2} \\[6pt]
						\cos\frac{\theta}{2} e^{-i\phi / 2}
					   \end{pmatrix}
						e^{i \zeta /2}, 
\end{align}
where $\theta \equiv \theta(t) \in [0,\pi]$, $\phi \equiv \phi(t) \in [-\pi,\pi]$, and $\zeta \equiv \zeta(t)$ denote the polar angle, azimuthal angle, and global phase, 
respectively. 

The solution of time-dependent Schr\"{o}dinger equation $i \frac{\partial}{\partial t} |\psi_\lambda(t)\rangle = H_\lambda^{\rm eff} |\psi_\lambda(t)\rangle$ is 
$|\psi_\lambda(t)\rangle$. Using perturbation theory up to $\mathcal{O}(\lambda^2)$, we have 
\begin{align}
	|\psi_\lambda(t_f)\rangle &\simeq |\psi^{(0)}(t_f)\rangle + |\psi^{(1)}(t_f)\rangle, \cr
	|\psi^{(1)}(t_f)\rangle &= - i \lambda \int_{0}^{t_f} U_0(t_f,t) H_1(t) |\psi^{(0)}(t)\rangle dt, 
\end{align}
where $|\psi^{(0)}(t_f)\rangle = |\psi(t_f)\rangle$ is the unperturbed solution, and the time evolution operator without perturbation is given by 
\begin{align}
	U_0(t_f,t) = |\psi(t_f)\rangle\langle\psi(t)|+|\psi_\bot(t_f)\rangle\langle\psi_\bot(t)|. 
\end{align}

The fidelity of the target state at the final time $t_f$ is defined as $F = |\langle\psi(t_f)|\psi_\lambda(t_f)\rangle|^2$, i.e., 
\begin{align}\label{P_err}
	F \approx 1 - \lambda^2 \left| \int_{0}^{t_f} \langle\psi_\bot(t)|H_1(t)|\psi(t)\rangle dt \right|^2. 
\end{align}
The influence of the errors in single-photon drive field can be evaluated by the error sensitivity. Defining the systematic error sensitivity as \cite{Ruschhaupt2012NJP}
\begin{align}
	q_{s,\lambda} = - \frac{1}{2} \frac{\partial^2 F}{\partial \lambda^2} \Bigg|_{\lambda=0}, 
\end{align}
we have $F \approx 1 - \lambda^2\,q_{s,\lambda}$ and 
\begin{align*}
	q_{s,\lambda} &= \left| \int_{0}^{t_f} \langle\psi_\bot(t)|H_1(t)|\psi(t)\rangle dt \right|^2 \cr
				 &= \frac{1}{4} \left| \int_{0}^{t_f} e^{-i\zeta} \left[\Omega_R \left(\cos^2\frac{\theta}{2}e^{i\phi} - \sin^2\frac{\theta}{2}e^{-i\phi}\right)\right] dt \right|^2. 
\end{align*}
It can be observed that the magnitude of the error sensitivity depends on the function of the control field and the evolution path. 

\subsection{Inverse engineering via Bloch vector trajectory}\label{Inverse engineering}
The evolution of the normalized state $|\psi(t)\rangle$ is governed by the time-dependent Schr\"{o}dinger equation 
$i \frac{\partial}{\partial t} |\psi(t)\rangle = H_0^{\rm eff} |\psi(t)\rangle$. 
To explicitly derive the control protocol via inverse engineering, we represent the system dynamics in the Bloch vector representation \cite{Berry2009,Chen2010PRL,Feynman1957}. 
The effective control Hamiltonian can then be parameterized as $H_0^{\rm eff} = \frac{1}{2} \boldsymbol{B} \cdot \boldsymbol{\sigma}$, thus 
\begin{align}\label{B}
	B_x = \Omega_R,\ \ \ B_y = 0,\ \ \ B_z = 0, 
\end{align}
where $B_k\ (k=x,y,z)$ represent the components of the effective magnetic field vector $\boldsymbol{B}$ along the $k$-direction of the Bloch sphere and $\boldsymbol{\sigma}$ is 
the vector of Pauli matrices. 

The Bloch vector $\boldsymbol{r} = \langle \boldsymbol{\sigma} \rangle$, therefore 
\begin{align}\label{r}
	r_x &= \langle\psi| \sigma_x |\psi\rangle = \sin\theta \cos\phi, \cr
	r_y &= \langle\psi| \sigma_y |\psi\rangle = -\sin\theta \sin\phi, \cr
	r_z &= \langle\psi| \sigma_z |\psi\rangle = \cos\theta. 
\end{align}
Here, $|\psi\rangle \equiv |\psi(t)\rangle$ in Eq.~(\ref{psi_t}). 

Substituting Eqs.~(\ref{B}) and (\ref{r}) into Bloch equations $\dot{\boldsymbol{r}} = \boldsymbol{B} \times \boldsymbol{r}$ \cite{Feynman1957}, we can obtain 
\begin{align}\label{angle}
	\dot{\phi} &= \Omega_R \cot\theta \cos\phi, \cr
	\dot{\theta} &= \Omega_R \sin\phi.
\end{align}
To obtain the global phase $\zeta$, starting from the time-dependent Schr\"{o}dinger equation, we have 
\begin{align}
	i \langle\psi\dot{|\psi\rangle} = \langle\psi| H_0^{\rm eff} |\psi\rangle, 
\end{align}
where $\dot{|\psi\rangle} \equiv \frac{\partial}{\partial t} |\psi(t)\rangle$, then we obtain 
\begin{align}\label{Phase}
	\dot{\zeta} &= 2\langle\psi| H_0^{\rm eff} |\psi\rangle + \dot{\phi} \cos\theta \cr
				&= \Omega_R \cos\phi / \sin\theta. 
\end{align}

Then, substituting Eqs.~(\ref{angle}) and (\ref{Phase}) into the systematic error sensitivity $q_{s,\lambda}$, we have 
\begin{align}\label{qs}
	q_{s,\lambda} &= \frac{1}{4} \left| \int_{0}^{t_f} e^{-i\zeta} \left[\Omega_R \left(\cos^2\frac{\theta}{2}e^{i\phi} - \sin^2\frac{\theta}{2}e^{-i\phi}\right)\right] dt \right|^2 \cr
				 &= \frac{1}{4} \left| \int_{0}^{t_f} e^{-i\zeta} (\dot{\zeta}\sin\theta\cos\theta + i \dot{\theta}) dt \right|^2. 
\end{align}

\subsection{Control protocol}\label{Nopt}
The system dynamics are described by 
\begin{align}\label{Dynamics}
	\dot{\rho}(t) = -i[H_\lambda(t),\rho(t)] + \kappa_2 \mathcal{D}[a^2-\alpha^2] \rho(t).
\end{align}
For simplicity, we assume an ideal situation without parameter imperfections, i.e. $\lambda=0$. 
To achieve the state transfer of cat-state from $|\mathcal{C}_+\rangle$ to $|\mathcal{C}_-\rangle$ following the evolution state in 
Eq.~(\ref{psi_t}), one needs to set the boundary conditions 
\begin{align}\label{Boundary}
	\theta(0) = 0,\ \ \ \ \theta(t_f) = \pi.
\end{align}
Here, $t_f$ denotes  the final time of the evolution. 
For verification, we can impose the conditions as follow 
\begin{align}\label{boundary}
	\dot{\theta}(0)&=\dot{\theta}(t_f)=0, 
\end{align}
and choose $\phi(t) \simeq \pi/2$ to minimize the effective Rabi frequency, imposing 
\begin{align}
	\phi(0) &= \phi(t_f/2) = \phi(t_f) = \pi/2, \cr
	\dot{\phi}(0) &= \pi/(10t_f),\ \dot{\phi}(t_f) = \pi/(10t_f). 
\end{align}
{It should be noted that $\dot{\phi} = \Omega_R \cot\theta \cos\phi$ may exhibit singular behavior when $\theta \rightarrow 0$ or $\theta \rightarrow \pi$, 
corresponding to the initial and final times of the state transfer protocol. To eliminate this singularity, we impose the boundary conditions $\phi(0) = \phi(t_f) = \pi/2$, 
which ensure that $\cos\phi = 0$ at the boundaries. As a result, the potentially divergent term remains finite and approaches zero as $\theta \rightarrow 0$ or 
$\theta \rightarrow \pi$.} 

\begin{figure}
	\centering
	\scalebox{0.2616}{\includegraphics{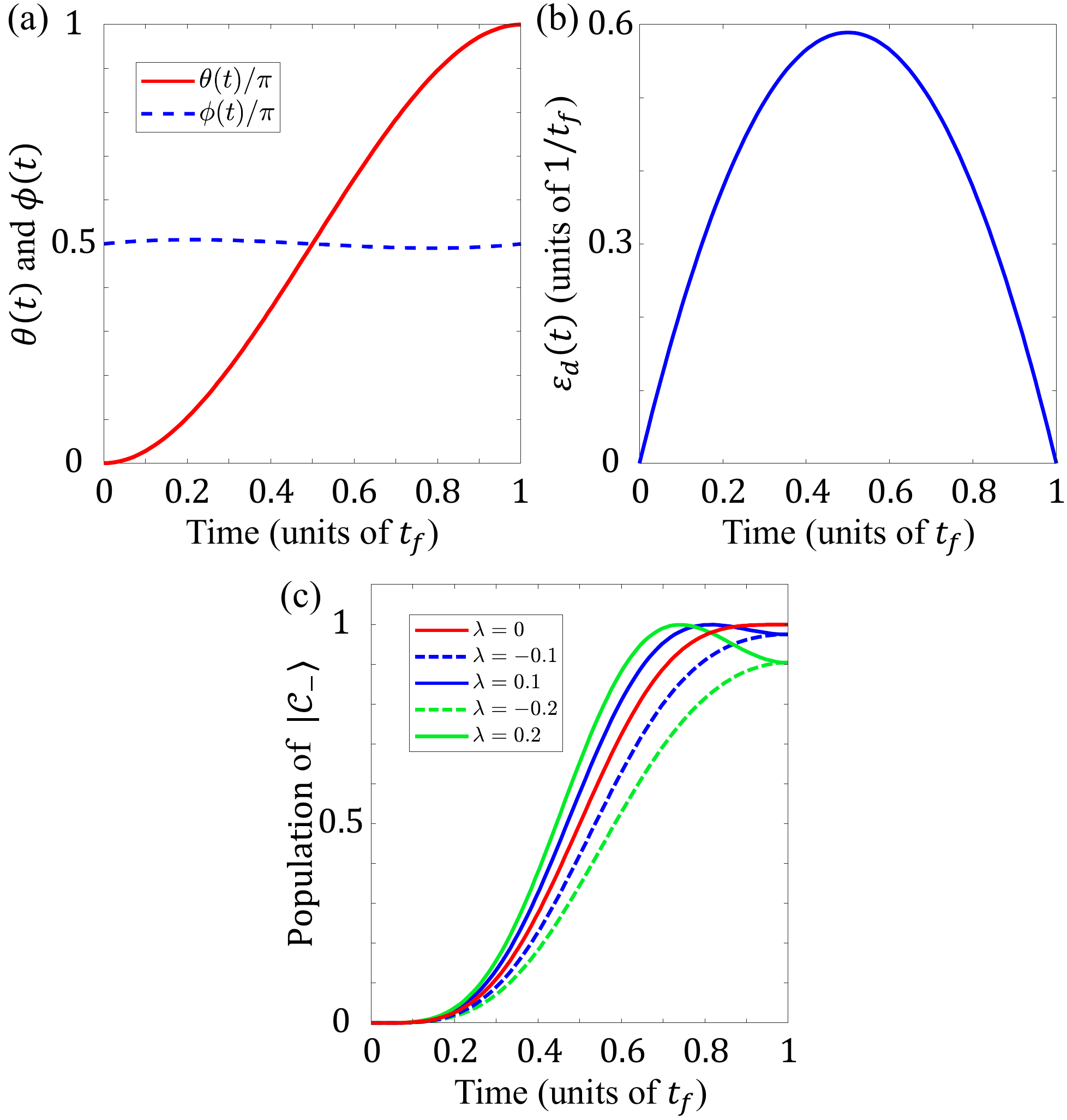}}
	\caption{(a) $\theta(t)$ and $\phi(t)$ as a function of dynamics evolution time. 
			{(b) Single photon driving strength $\varepsilon_d(t)$ as a function of dynamics evolution time.} 
			(c) State transfer from $|\mathcal{C}_+\rangle$ to $|\mathcal{C}_-\rangle$. The populations of the odd cat state $|\mathcal{C}_-\rangle$ are defined as 
			$P_- = |\langle\mathcal{C}_-|\rho_a|\mathcal{C}_-\rangle|$, where $\rho_a = {\rm Tr}_b[\rho]$ determined by Eq.~(\ref{Dynamics}). 
			Here, we assume $\kappa_a = 1\ {\rm kHz}$, $\kappa_b = 5\ {\rm MHz}$, and $g = 1\ {\rm MHz}$ that $\kappa_2 = 4g^2/\kappa_b = 0.8\ {\rm MHz}$. 
			The evolution time $t_f = 0.1/\kappa_a$. }
	\label{Verification}
\end{figure}

Following the above boundaries, we can assume polynomial forms $\theta(t) = \sum_{i=0} a_i t^i$, and $\phi(t) = \sum_{i=0} b_i t^i$ that lead to 
\begin{align}
	\theta(t) &= \frac{\pi}{t_f^3} (3t_f t^2 - 2t^3), \cr
	\phi(t) &= \frac{\pi}{10t_f^3} (2t^3 - 3t_f t^2 + t_f^2 t + 5t_f^3). 
\end{align}
{The control field is constructed through an inverse engineering approach based on the auxiliary Eq.~(\ref{angle}), which may exhibit singularities owing to 
the cotangent term involved in the equation \cite{Martinez2013PRL}. Therefore, it is essential to ensure the smoothness of the resulting control field. 
As shown in Fig.~\ref{Verification}(b), the single-photon drive strength $\varepsilon_d(t)$ varies smoothly throughout the entire evolution.} 

When there are no parameter imperfections, as shown by the red-solid curve in Fig.~\ref{Verification}(c), such parameters enable a high-fidelity state transfer from 
the even cat state $|\mathcal{C}_+\rangle$ to the odd cat state $|\mathcal{C}_-\rangle$. However, this control protocol has poor robustness. When parameter imperfections are 
present, i.e., $\lambda \neq 0$, the final population of $|\mathcal{C}_-\rangle$ decreases as the magnitude of the error rate $|\lambda|$ increases.

{Although these systematic errors can induce undesired bit-flip errors during the population transfer process, they can generally be mitigated through proper 
calibration or the use of improved hardware once their origins are identified \cite{Krantz2019APR}. However, in practice, it is often difficult to identify all the 
underlying error mechanisms because systematic errors can arise from multiple experimental factors (as briefly discussed in Sec.~\ref{Hc}). As a result, residual control 
errors are usually unavoidable. In modern superconducting platforms, the relative control-amplitude error can still reach up to $10\%$ in the absence of continuous 
recalibration \cite{Rol2019PRL}. Therefore, designing a control protocol that is inherently robust against systematic errors is highly desirable.}

\section{Optimal robust population transfer}\label{Opt}
\subsection{Reducing error sensitivity}
In order to enhance the robustness of the control protocol, one needs to set the systematic error sensitivity $q_{s,\lambda} = 0$. Defining the global phase $\zeta(t)$ to be a 
function of $\theta$, i.e., $\widetilde{\zeta}(\theta) \equiv \zeta(t)$, the systematic error sensitivity in Eq.~(\ref{qs}) becomes 
\begin{align}
	q_{s,\lambda} = \frac{1}{4} \left| \int_{\theta_0}^{\theta_f} e^{-i\widetilde{\zeta}} (\dot{\widetilde{\zeta}}\cos\theta\sin\theta + i) d\theta \right|^2, 
\end{align}
where $\theta_0 \equiv \theta(0)$, $\theta_f \equiv \theta(t_f)$, and $\dot{\widetilde{\zeta}} \equiv d\widetilde{\zeta} / d\theta = \dot{\zeta} / \dot{\theta}$. 

For simplicity, we assume $\widetilde{\zeta}(\theta) = m\theta + \zeta_0$ that $\dot{\widetilde{\zeta}}=m$, where $\zeta_0$ is a constant. Then we obtain 
\begin{align*}
	q_{s,\lambda} &= \frac{1}{4} \left| \int_{\theta_0}^{\theta_f} e^{-i m\theta} (m\cos\theta\sin\theta + i) d\theta \right|^2 \cr
		&= \frac{1}{4} \left| \frac{m}{2} \int_{\theta_0}^{\theta_f} e^{-i m\theta} \sin2\theta d\theta + i\int_{\theta_0}^{\theta_f} e^{-i m\theta} d\theta \right|^2. 
\end{align*}
It can be calculated that $m=4$ satisfies $q_{s,\lambda}=0$ (more details in Appendix~\ref{Apd_C}). According to Eq.~(\ref{Phase}), we have 
$\dot{\widetilde{\zeta}} = \dot{\zeta} / \dot{\theta} = \cot\phi/\sin\theta = 4$, thus 
\begin{align}\label{Eq31}
	\cot\phi = 4\sin\theta. 
\end{align}

Taking the time derivative of both sides of Eq.~(\ref{Eq31}), one obtains 
\begin{align}\label{Eq32}
	\dot{\phi} &= -4 \dot{\theta} \cos\theta \sin^2\phi. 
\end{align}
Substituting Eq.~(\ref{Eq31}) into Eq.~(\ref{angle}) yields
\begin{align}\label{Eq33}
	\dot{\phi} &= 4 \dot{\theta} \cos\theta. 
\end{align}
It can be observed that Eq.~(\ref{Eq32}) and Eq.~(\ref{Eq33}) are mutually contradictory, i.e., Eq.~(\ref{angle}) and Eq.~(\ref{Eq31}) cannot be satisfied simultaneously. 
Therefore, an additional control field needs to be introduced.

\subsection{Additional control field}
The effective control Hamiltonian in Eq.~(\ref{H0}) contains only the Rabi oscillation term, i.e., $\frac{1}{2} \Omega_R \sigma_x$. 
In order to introduce an additional control degree of freedom, we further consider adding an effective detuning term, i.e., $\frac{1}{2} \Delta \sigma_z$. 

To realize this additional control freedom, we employ the following Hamiltonian in the interaction picture \cite{Cohen2017PRL}: 
\begin{align}\label{Eq10}
	H_{\rm int}(t) &= -E'_J(t) \cos[\varphi_a(a e^{-i\omega_a t} + a^\dagger e^{i\omega_a t})] \cr
		   		   &= -\frac{E'_J}{2} [ D(i\varphi_a e^{i\omega_a t}) + D^\dagger(i\varphi_a e^{i\omega_a t}) ], 
\end{align}
where $E'_J$ is the effective Josephson energy, $\omega_a$ is the frequency of the storage cavity, and $\varphi_a = \sqrt{Z_a/2R_Q}$ is the dimensionless phase,  with $Z_a$ 
and $R_Q$ denoting the impedance of the cavity mode seen by the junction and the superconducting resistance quantum, respectively. In the context of quantum superconducting 
circuits, the interaction Hamiltonian in Eq.~(\ref{Eq10}) can be realized by strongly coupling a high impedance cavity to a Josephson junction and assuming that other modes 
(including the junction mode) are never excited \cite{Koch2007PRA,Masluk2012PRL,Pop2014,Cohen2017PRL}. 

When the condition $E'_J \ll \omega_a$ is satisfied, the interaction Hamiltonian under the RWA can be calculated as 
\begin{align}
	H_{\rm int}^{\rm RWA}(t) = E'_J(t) e^{-\varphi_a^2/2} \sum_{m=0}^{\infty} L_m(\varphi_a^2)|m\rangle\langle m|, 
\end{align}
where $L_m(*)$ is the Laguerre polynomial of order $m$. Projecting onto the cat-state subspace $\left\{|\mathcal{C}_{\pm}\rangle\right\}$, the effective Hamiltonian 
becomes \cite{Cohen2017PRL}
\begin{align}
	H_{\rm int}^{\rm eff}(t) &= \frac{E'_J(t) e^{-(\varphi_a - 2\alpha)^2 / 2}}{-2\sqrt{\pi \alpha \varphi_a}} \sigma_z + \mathcal{O}(E'_J e^{-\varphi_a^2 /2}) \cr
							 &\simeq \frac{1}{2} \Delta(t) \sigma_z, 
\end{align}
under the condition $\varphi_a=2\alpha$, the effective detuning becomes $\Delta(t) = -E'_J(t) / \alpha \sqrt{2\pi}$,

Now, the control Hamiltonian $H_c(t) = H_{\rm int}^{\rm RWA}(t) + H_0(t)$ in the cat-state subspace becomes  
\begin{align}\label{H_c}
	H_c^{\rm eff} = \frac{1}{2}
		   \begin{pmatrix}
			\Delta& \Omega_R \\[6pt]
			\Omega_R& -\Delta
		   \end{pmatrix}, 
\end{align}
where $\Delta \equiv \Delta(t)$, and $\Omega_R \equiv \Omega_R(t)$. 

Following Sec.~\ref{Inverse engineering}, we can obtain 
\begin{align}
	\dot{\phi} &= \dot{\theta} \cot\theta \cot\phi - \Delta, \cr
	\dot{\theta} &= \Omega_R \sin\phi, \cr
	\dot{\zeta} &= \dot{\theta} \cot\phi/\sin\theta. 
\end{align}
Therefore, we have 
\begin{align}\label{Eq34}
	\Delta &= \dot{\zeta}\cos\theta - \dot{\phi}, \cr
	\Omega_R &= \dot{\theta} / \sin \phi. 
\end{align}
The optimal robustness condition is 
\begin{align}\label{Op}
	\cot\phi = 4\sin\theta. 
\end{align}

\begin{figure*}
	\centering
	\scalebox{0.58}{\includegraphics{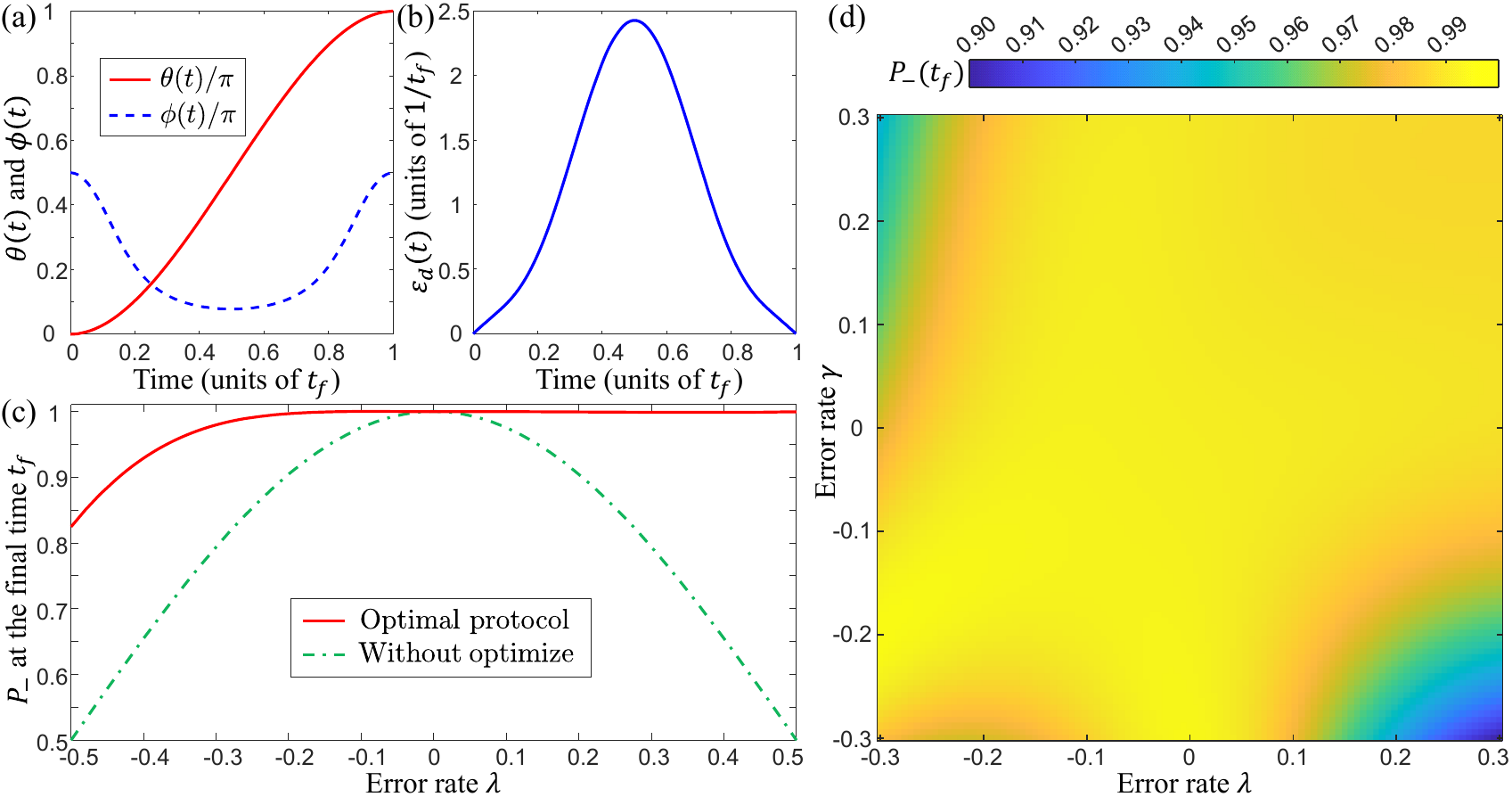}} 
	\caption{(a) $\theta(t)$ and $\phi(t)$ as a function of dynamics evolution time. 
			(b) Single photon driving strength $\varepsilon_d(t)$ as a function of dynamics evolution time. 
			(c) Population of the odd cat-state $|\mathcal{C}_-\rangle$ at the final time $t_f$ versus parameter imperfection in $\varepsilon_d$ (with error rate $\lambda$). 
			The green dashed-dotted curve denotes the protocol without optimize in Sec.~\ref{Nopt}, while the red solid curve denotes the results of the optimal protocol in 
			Sec.~\ref{Opt}. 
			(d) Population of the odd cat-state $|\mathcal{C}_-\rangle$ at the final time $t_f$ versus parameter imperfection in $E'_J$ (with error rate $\gamma$) and 
			$\varepsilon_d$ (with error rate $\lambda$). Parameters are the same as those in Fig.~\ref{Verification}. }
	\label{Optimal}
\end{figure*}

Considering the above optimal robustness condition and the boundary condition in Eq.~(\ref{Boundary}), we can redesign the parameters $\theta(t)$ and $\phi(t)$, as shown in 
Fig.~\ref{Optimal}(a). Using this optimal protocol, the control scheme becomes insensitive to the deviation on the single-photon drive $\varepsilon_d$. 
As shown in Fig.~\ref{Optimal}(c), when a parameter imperfection rate in single-photon drive $\varepsilon_d$ is $\pm20\%$, the population of the odd cat-state 
$|\mathcal{C}_-\rangle$ at the final time $P_-(t_f) > 99.5\%$, resulting in an optimally robust state transfer. 

Then, we further consider that the presence of parameter imperfections in $E'_J$, characterized by the error rate $\gamma$. 
{ In experiments, these imperfections mainly originate from magnetic-flux crosstalk between neighboring control lines and fluctuations in the flux-bias 
source \cite{Lescanne2020NP}. Although such systematic errors can be mitigated through careful calibration, they are generally difficult to eliminate completely 
because of the complexity of the control circuitry. In modern superconducting platforms, despite careful calibration, relative flux-crosstalk errors on the order of 
several percent to $10\%$ typically remain in multi-mode architectures \cite{Abrams2019PRAppl}. As shown in Fig.~\ref{Optimal}(d), even with a relative parameter 
error $\gamma$ of up to $10\%$, the control protocol remains robust and achieves high-fidelity state transfer.} 

\section{Decoherence}
While the influence of the pure dephasing and single-photon loss was assumed to be negligible in the preceding derivations to simplify the control protocol design, these noise 
sources are unavoidable in realistic experimental platforms. In this section, we take these two types of noise into consideration. The Lindblad master equation becomes 
\begin{align}\label{Decoherence}
	\dot{\rho}(t) =& -i[H_{\rm tot}(t),\rho(t)] + \kappa_2 \mathcal{D}[a^2-\alpha^2] \rho(t) \cr
				   &+ \kappa_a^\phi \mathcal{D}[a^\dagger a] \rho(t) + \kappa_a \mathcal{D}[a] \rho(t). 
\end{align}

\subsection{Pure dephasing}\label{Pure dephasing}
Considering only the pure dephasing $\kappa_a^\phi \mathcal{D}[a^\dagger a] \rho(t)$. {In high-$Q$ cavities, pure dephasing is often attributed to random 
fluctuations in the dispersive frequency shift induced by residual photon-number fluctuations in auxiliary modes, such as the readout resonator or the nonlinear coupler. 
These fluctuations result in stochastic phase drift of the cat state stored in the cavity and eventually destroy its phase coherence \cite{Sears2012PRB}. In modern 
superconducting platforms, the pure dephasing rate can typically be suppressed to a level below the single-photon loss rate \cite{Lescanne2020NP, Gautier2023}.} 

The pure dephasing can cause leakage out of the cat-state subspace \cite{Puri2020,Puri2019PRX}, since 
\begin{align}
	a^\dagger a |\mathcal{C}_\pm\rangle = |\alpha|^2 |\mathcal{C}_\pm\rangle + \alpha \frac{N_\pm}{N_{\mp}^{e,1}} |\psi_{\mp}^{e,1}\rangle. 
\end{align}
Considering the full Hilbert space, the projection operator is modified to 
\begin{align}
	P = |\mathcal{C}_+\rangle \langle\mathcal{C}_+| + |\mathcal{C}_-\rangle \langle\mathcal{C}_-| + \sum_{n=1}^{\infty} |\psi_\pm^{e,n}\rangle \langle\psi_\pm^{e,n}|.
\end{align}
Therefore, the term describing pure dephasing becomes 
$\kappa_a^\phi \mathcal{D}[a^\dagger a] \rho(t) \Longrightarrow \kappa_a^\phi \mathcal{D}[P a^\dagger a P] \rho(t)$, i.e., 
\begin{align}\label{L2}
	&\ \kappa_a^\phi \mathcal{D}[P a^\dagger a P] \rho(t) \cr
	= &\ \kappa_a^\phi |\alpha|^4 \mathcal{D}[|\mathcal{C}_+\rangle \langle\mathcal{C}_+| + |\mathcal{C}_-\rangle \langle\mathcal{C}_-|] \rho(t) \cr
	 &+ \kappa_a^\phi |\alpha|^2 \mathcal{D}[\mathcal{N}_+ |\psi_{-}^{e,1}\rangle \langle\mathcal{C}_+| + \mathcal{N}_- |\psi_{+}^{e,1}\rangle \langle\mathcal{C}_-|] \rho(t) \cr
	 &+ \cdots, 
\end{align}
where $\mathcal{N}_+ = {N_+}/{N_{-}^{e,1}}$ and $\mathcal{N}_- = {N_-}/{N_{+}^{e,1}}$ are the dimensionless constant. In the limit of large $\alpha$, 
$\mathcal{N}_+ \approx \mathcal{N}_- \approx 1$. According to Eq.~(\ref{L2}), pure dephasing does not cause bit-flip errors, but it can cause transitions from cat-states 
$|\mathcal{C}_\pm\rangle$ to the first-excited states $|\psi_\mp^{e,1}\rangle$, with a leakage rate $\Gamma_{\rm Leak} \approx \kappa_a^\phi |\alpha|^2$. 

\begin{figure}
	\centering
	\scalebox{0.46}{\includegraphics{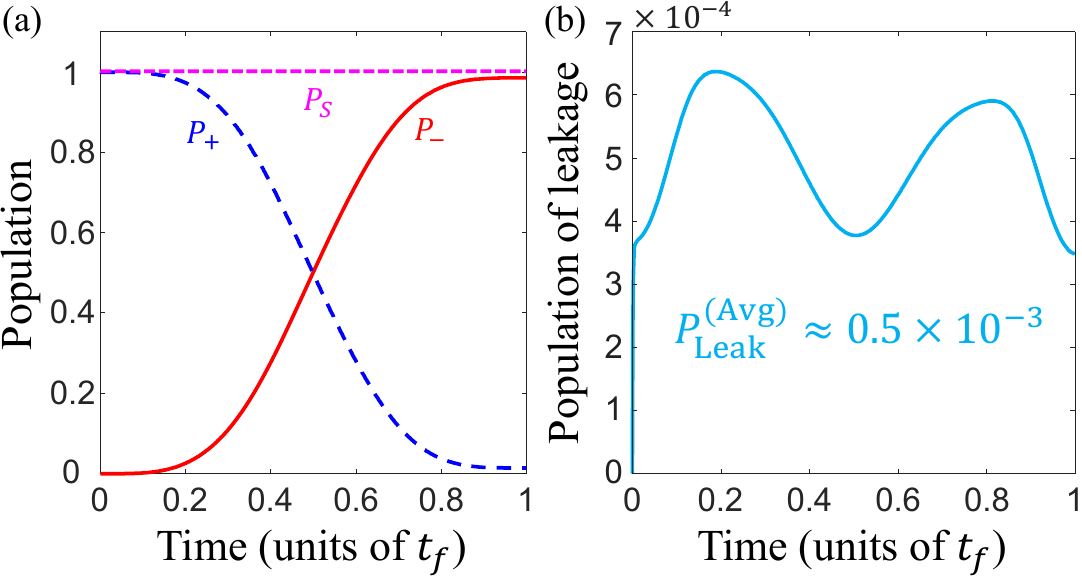}}
	\caption{(a) State transfer from $|\mathcal{C}_+\rangle$ to $|\mathcal{C}_-\rangle$ with pure dephasing. 
			 (b) The population of leakage out of cat-state subspace. Parameters are the same as those in Fig.~\ref{Verification}. }
	\label{Pure}
\end{figure}

Noting that, the two-photon dissipation confinement $\kappa_2 \mathcal{D}[a^2-\alpha^2] \rho(t)$ can force any states toward cat-state subspace 
$\left\{|\mathcal{C}_{\pm}\rangle\right\}$ \cite{Mirrahimi2014NJP,Leghtas2015,Reglade2024}. Projecting the dissipator onto the cat-state subspace, 
we can obtain 
\begin{align}\label{TpD}
	&\ \kappa_2 \mathcal{D}[P(a^2-\alpha^2)P] \rho(t) \cr
	\approx&\ 4\kappa_2 |\alpha|^2 \mathcal{D}[|\mathcal{C}_+\rangle \langle\psi_{-}^{e,1}| + |\mathcal{C}_-\rangle \langle\psi_{+}^{e,1}|] \rho(t) \cr
	 &+ \cdots. 
\end{align}
Here, we focus on the dissipative stabilization process of the first excited state. 
According to Eq.~(\ref{TpD}), the two-photon dissipation confinement can force the first-excited states $|\psi_\mp^{e,1}\rangle$ toward cat-states $|\mathcal{C}_\pm\rangle$ 
with a restoration rate $\Gamma_{\rm R,1} \approx 4\kappa_2 |\alpha|^2$. 

Without loss of generality, the pure dephasing rate $\kappa_a^\phi \lesssim \kappa_a$ is much smaller than the two-photon dissipation rate $\kappa_2$ that satisfies 
$\Gamma_{\rm Leak} \ll \Gamma_{\rm R,1}$. 
Even though pure dephasing $\kappa_a^\phi \mathcal{D}[a^\dagger a] \rho(t)$ tends to drive transitions from the 
logical states $|\mathcal{C}_\pm\rangle$ toward the first-excited states $|\psi_\mp^{e,1}\rangle$, the two-photon dissipation provides an autonomous restoration force that 
rapidly pumps these excited states back into the steady-state subspace. Thus, the population of leakage out of cat-state subspace $\left\{|\mathcal{C}_{\pm}\rangle\right\}$ 
scales as $\mathcal{O}(\Gamma_{\rm Leak} / \Gamma_{\rm R,1})$, the leakage caused by pure dephasing can be effectively suppressed. 

\begin{figure}
	\centering
	\scalebox{0.713}{\includegraphics{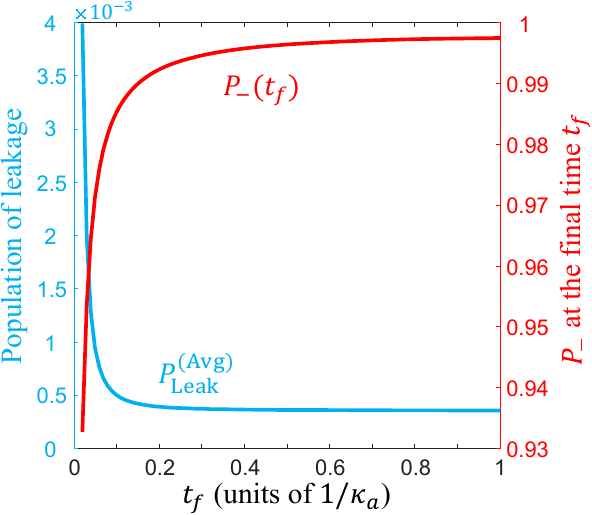}}
	\caption{With pure dephasing, the population of $|\mathcal{C}_-\rangle$ at the final time $P_-(t_f)$ is denoted by the red solid curve. The light-blue solid curve denotes 
			 the average of the population of leakage. The evolution time $t_f \in (0,1/\kappa_a]$ and other parameters are the same as those in Fig.~\ref{Verification}. }
	\label{Pure2}
\end{figure}

We can define the population of leakage out of cat-state subspace $\left\{|\mathcal{C}_{\pm}\rangle\right\}$ as $P_{\rm Leak} = 1 - P_S$, 
where $P_S = P_+ + P_-$ denotes the total populations within the cat-state subspace $\left\{|\mathcal{C}_{\pm}\rangle\right\}$. 
The average of the population of leakage is defined as follow
\begin{align}
	P_{\rm Leak}^{(\rm Avg)} = \frac{1}{t_f} \int_{0}^{t_f} P_{\rm Leak}(t) dt. 
\end{align}

As shown in Fig.~\ref{Pure}(a), assuming the pure dephasing rate $\kappa_a^\phi = \kappa_a$, the total populations within the cat-state subspace 
$\left\{|\mathcal{C}_{\pm}\rangle\right\}$ does not reduce. The population of leakage out of cat-state subspace $\left\{|\mathcal{C}_{\pm}\rangle\right\}$, 
as shown in Fig.~\ref{Pure}(b), has been suppressed by two-photon dissipation confinement. 

Figure \ref{Pure}(a) shows that, although the leakage caused by pure dephasing can be well suppressed, two-photon dissipation confinement results in bit-flip errors, the 
population of $|\mathcal{C}_-\rangle$ at the final time $P_-(t_f) \approx 98.53\%$. As shown by the red solid curve in Fig.~\ref{Pure2}, bit-flip errors can be reduced by 
extending the evolution time $t_f$. When $t_f = 0.5/\kappa_a$, the population $P_-(t_f) \approx 99.64\%$. The light-blue curve in Fig.~\ref{Pure2} denotes the average of the 
population of leakage, it further demonstrates that the leakage caused by pure dephasing can be effectively suppressed.

\subsection{Single-photon loss}
All the considerations made in the previous text were based on the absence of damping in storage mode. Now we consider single-photon loss $\kappa_a \mathcal{D}[a] \rho(t)$, 
{it is the dominant decoherence channel for bosonic codes and can arise from various loss mechanisms, including dielectric and conductor losses in the cavity as 
well as the Purcell effect through the readout port} \cite{Wang2015APL, Sete2015PRA}.

Within the cat-state subspace $\left\{|\mathcal{C}_{\pm}\rangle\right\}$, the term describing single-photon loss in the master equation for large $\alpha$ becomes 
\begin{align}\label{L1}
	\kappa_a \mathcal{D}[a] \rho(t) \Longrightarrow\ &\kappa_a \mathcal{D}[P_\mathcal{C} a P_\mathcal{C}] \rho(t) \cr
									\approx\ &\kappa_a |\alpha|^2 \mathcal{D}[\sigma_x] \rho(t). 
\end{align}
According to Eq.~(\ref{L1}), the single-photon dissipation does not lead to leakage out of the cat-state subspace, but it can only cause bit-flip errors. 

\begin{figure}
	\centering
	\scalebox{0.75}{\includegraphics{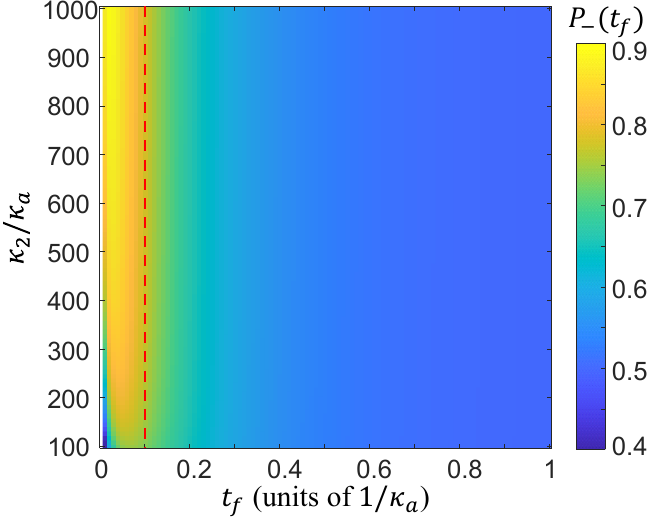}}
	\caption{Population of the odd cat-state $|\mathcal{C}_-\rangle$ at the final time $t_f$ versus two-photon dissipation rate $\kappa_2$ (units of $\kappa_a$) and 
			dynamics evolution time $t_f$ (units of $1/\kappa_a$). }
	\label{Sploss}
\end{figure}

As illustrated in Fig.~\ref{Sploss}, for a short evolution time $t_f \lesssim 0.1/\kappa_a$, single-photon dissipation mainly induces bit-flip errors, resulting in a final 
population of the odd cat state $P_-(t_f) \gtrsim 80\%$. However, when the evolution time $t_f$ is very short, achieving the target dynamics requires excessively large control 
Hamiltonian amplitudes according to Eq.~(\ref{Eq34}). 
For a finite two-photon dissipation rate $\kappa_2$, these large control terms break down the confinement imposed by the two-photon dissipation, leading to leakage out of the 
cat-state subspace $\left\{|\mathcal{C}_{\pm}\rangle\right\}$. In contrast, for long-time evolution, bit-flip errors due to damping in the storage cavity become increasingly 
significant. Ultimately, these competing effects lead to an unavoidable trade-off between leakage and decoherence. 
{Consequently, operating within a bounded time window, $\frac{1}{2\kappa_2 |\alpha|^3} \ll t_f \ll \frac{1}{\kappa_a |\alpha|^2}$, becomes essential to balance the 
preservation of dissipative confinement against the accumulation of decoherence (see Appendix~\ref{TW} for more details). }

{Single-photon loss is a dominant logical error channels in cat-state qubits. Although its influence can be reduced through an appropriate choice of the control 
parameters $\theta(t)$ and $\phi(t)$, it cannot be completely eliminated (see Appendix~\ref{qn} for more details). This is because single-photon loss originates from the 
intrinsic dissipative channels of the system and is therefore unavoidable in realistic experiments \cite{Gautier2023, Gao2021}. 
Moreover, single-photon loss is an irreversible non-unitary process, whereas control optimization is fundamentally based on engineering the unitary evolution of the system. 
Therefore, the effect of single-photon loss can only be reduced rather than completely eliminated. Fortunately, although single-photon loss cannot be completely eliminated 
through control optimization, its influence can be substantially reduced by improving the quality factor of the storage cavity. State-of-the-art superconducting cavities 
have already demonstrated single-photon lifetimes ranging from the millisecond to the second timescale \cite{Romanenko2020, Reagor2013APL, Chakram2021PRL}, with lifetimes 
reaching up to $2$ seconds in recent experiments \cite{Romanenko2020}. Therefore, continued advances in cavity design, materials, and fabrication techniques are expected 
to further suppress the impact of single-photon loss in future experimental implementations.}

\section{Discussion and conclusions}
The results presented in this manuscript demonstrate that dissipative confinement provides a natural and effective framework for robust control of cat-state qubits. By 
restricting the dynamics to the cat-state manifold, an effective two-level description becomes valid, allowing systematic robustness optimization to be carried out 
analytically. In particular, the engineered two-photon dissipation not only stabilizes the logical subspace but also strongly suppresses leakage induced by the pure dephasing, 
highlighting an intrinsic advantage of dissipative cat qubits over purely Hamiltonian schemes. At the same time, our analysis reveals a clear trade-off between control speed 
and dissipation-induced errors, which sets practical constraints on pulse amplitudes and operation times. These features provide useful physical guidance for the design of 
high-fidelity control protocols in realistic experimental settings. 

In conclusion, we have proposed an optimally robust STA control protocol for dissipative-cat qubits. By inverse engineering the system dynamics within the cat-state subspace, 
we achieve fast and high-fidelity state transfer that is insensitive to systematic imperfections. Lindblad simulations confirm the robustness of the protocol in the 
presence of decoherence. Our work establishes a clear connection between robustness optimization and dissipative confinement, and offers a practical route 
toward reliable control of bosonic qubits in current superconducting quantum circuits. 

\begin{acknowledgments}
S.-W.X. and Z.-Y.Z. were supported by Student Research Training Program (SRTP) of China under Grant No. 202510386053. 
Y.-T.S. was supported by Student Research Training Program (SRTP) of China under Grant No. 202510386052. 
Y.-H.C. was supported by the National Natural Science Foundation of China under Grant No. 12304390 and 12574386, the National Postdoctoral 
Overseas Talent Recruitment Program of China, the Fujian 100 Talents Program, and the Fujian Minjiang Scholar Program.
Y.X. was supported by the National Natural Science Foundation of China under Grant No. 62471143, the Key Program of National Natural Science 
Foundation of Fujian Province under Grant No. 2024J02008, and the project from Fuzhou University under Grant No. JG2020001-2.
\end{acknowledgments}

\section*{Author contributions}
Y.-H.C. conceived and developed the idea. 
S.-W.X., Z.-Y.Z., and Y.-T.S. analyzed the data and performed the numerical simulations, with help from Y.-H.C. and Y.X.. 
S.-W.X., Z.-Y.Z., and Y.-T.S. cowrote the paper with feedback from all authors.

\section*{Data availability}
The data used for obtaining the presented numerical results as well as for generating the plots is available on request. 
Please refer to yehong.chen@fzu.edu.cn

\section*{Competing interests}
The authors declare that they have no competing interests.

\appendix
\section{System Hamiltonian}
We consider a schematic setup represented in Fig.~\ref{Sch}(a). A nonlinear coupling between the modes of two cavities is achieved by linking them with a small transmission line 
containing a Josephson junction. The Hamiltonian of this device reads \cite{Mirrahimi2014NJP,Leghtas2015}
\begin{align}
	H_0 &= \omega_a a^\dagger a + \omega_b b^\dagger b - E_J ( \cos\varPhi + \varPhi^2 / 2 ), \cr
	\varPhi &= \varPhi_a (a + a^\dagger) + \varPhi_b (b + b^\dagger), 
\end{align}
where $E_J$ is the Josephson energy, $\varPhi_a$ ($\varPhi_b$) is the standard deviation of the zero point flux fluctuation for storage (readout) mode of frequency $\omega_a$ 
($\omega_b$). Here, we focus solely on the dynamics of the fundamental modes of the two cavities, assuming all other modes remain unexcited. 

Assuming $\varPhi \ll 1$ that we can neglect sixth and higher order terms in the expansion of the cosine . The readout mode have two drive fields: a weak resonant drive 
$\varepsilon_b(t)$ and a strong off-resonant pump $\varepsilon_p(t)$. The frequencies of storage and readout mode are shifted by the nonlinear coupling. 
The dressed frequencies are noted $\widetilde{\omega}_a$ and $\widetilde{\omega}_b$. We take 
\begin{align*}
	\varepsilon_b(t) &= 2 \varepsilon_b \cos(\widetilde{\omega}_b t), \cr
	\varepsilon_p(t) &= 2 \varepsilon_p \cos(\omega_p t), 
\end{align*}
with 
\begin{align}
	\omega_p = 2\widetilde{\omega}_a - \widetilde{\omega}_b. 
\end{align}
Then we place ourselves in a regime where rotating terms can be neglected. Under the rotating-wave approximation, the system Hamiltonian is given by \cite{Mirrahimi2014NJP,Leghtas2015}
\begin{align}\label{A3}
	H_{ab} =&\ g_2^* a^2 b^\dagger + g_2 a^{\dagger 2} b - \varepsilon_b^* b - \varepsilon_b b^\dagger \cr
			& + \frac{\chi_{aa}}{2} (a^\dagger a)^2 + \frac{\chi_{bb}}{2} (b^\dagger b)^2 \cr
			& + \chi_{ab} (a^\dagger a)(b^\dagger b). 
\end{align}
When the induced cross-Kerr and self-Kerr terms, $\chi_{ab}$ and $\chi_{aa}$, $\chi_{bb}$, can be reduced from the Hamiltonian in Eq.(\ref{A3}), we can find 
\begin{align}
	g_2 = \frac{\varepsilon_p \chi_{ab}}{2(\omega_p-\widetilde{\omega}_b)}, 
\end{align}
denotes the rate of photon pairs are disspated from the storage cavity to single photons in the readout cavity, and then dissipated into the environment.

\section{Adiabatically eliminate the readout mode}\label{Eliminate}
We can obtain a master equation for the reduced density matrix of the storage mode by adiabatically eliminate the readout mode. The system dynamics are governed by the Lindblad 
master equation 
\begin{align}\label{Rho}
	\dot{\rho} = &-i[H_{ab},\rho] + \kappa_a \mathcal{D}[a]\rho + \kappa_b \mathcal{D}[b]\rho \cr
					&+ \kappa_a^\phi \mathcal{D}[a^\dagger a]\rho + \kappa_b^\phi \mathcal{D}[b^\dagger b]\rho, 
\end{align}
where $\rho$ is the joint density matrix. 

In the regime $\kappa_a^\phi,\ \kappa_b^\phi \ll \kappa_b$, we temporarily disregard the influence of the pure dephasing. Assuming that the number of photons in the readout 
mode is always much smaller than one. Therefore, we can search for a solution of Eq.~(\ref{Rho}) in the form 
\begin{align}\label{B2}
	\rho =&\ \rho_{00}|0\rangle\langle0| + \delta(\rho_{01}|0\rangle\langle1|+\rho_{10}|1\rangle\langle0|) \cr
			   & + \delta^2(\rho_{11}|1\rangle\langle1|+\rho_{02}|0\rangle\langle2|+\rho_{20}|2\rangle\langle0|) \cr
			   & + \mathcal{O}(\delta^3), 
\end{align}
where $\rho_{mn}$ acts on the storage mode Hilbert space, $|m\rangle\langle n|$ acts on the readout mode Hilbert space, and $\delta \ll 1$ is a dimensionless parameter. 
We place ourselves in the regime where $g/\kappa_b,\ \varepsilon_b/\kappa_b \sim \delta$ and $\kappa_a/\kappa_b \sim \delta^2$. 

Our goal is to derive the dynamics of the reduced density matrix of the storage mode by tracing out the readout mode, 
\begin{align}
	\rho_a = {\rm Tr}_b[\rho] \approx \rho_{00} + \delta^2 \rho_{11}. 
\end{align}

First, lets multioly Eq.~(\ref{B2}) by $\langle0|$ and $|0\rangle$. Then we obtain 
\begin{align}\label{B4}
	\dot{\rho}_{00} =& -i \delta^2 (\mathcal{A}^\dagger \rho_{10} - \rho_{01}\mathcal{A}) + \kappa_a \mathcal{D}[a]\rho_{00} \cr
					 & +\delta^2 \kappa_b \rho_{11} + \mathcal{O}(\delta^3), 
\end{align}
where $\mathcal{A} = (g a^2 - \varepsilon_b) / \delta$. 

Now, we need to find expressions of $\rho_{01}$, $\rho_{10}$, and $\rho_{11}$ up to $0^{th}$ order terms in $\delta$. Therefore, we find 
\begin{align}
	\dot{\rho}_{01} &= i \rho_{00} \mathcal{A}^\dagger - \frac{\kappa_b}{2} \rho_{01} + \mathcal{O}(\delta), \cr
	\dot{\rho}_{10} &= -i \mathcal{A} \rho_{00} - \frac{\kappa_b}{2} \rho_{10} + \mathcal{O}(\delta), \cr
	\dot{\rho}_{11} &= -i(\mathcal{A}\rho_{01}-\rho_{10}\mathcal{A}^\dagger) - \kappa_b \rho_{11} + \mathcal{O}(\delta).
\end{align}
Then we make the adiabatic approximation \cite{Mirrahimi2014NJP,Leghtas2015}, assuming that $\rho_{01}$, $\rho_{10}$, and $\rho_{11}$ are continuously in steady state, i.e., 
\begin{align*}
	\dot{\rho}_{01} \approx 0,\ \dot{\rho}_{10} \approx 0,\ {\rm and}\ \dot{\rho}_{11} \approx 0. 
\end{align*}
Therefore, we have 
\begin{align}
	\rho_{01} &\approx 2i \rho_{00} \mathcal{A}^\dagger / \kappa_b + \mathcal{O}(\delta), \cr
	\rho_{10} &\approx -2i \mathcal{A} \rho_{00} / \kappa_b + \mathcal{O}(\delta), \cr
	\rho_{11} &\approx -i (\mathcal{A}\rho_{01}-\rho_{10}\mathcal{A}^\dagger) / \kappa_b + \mathcal{O}(\delta) \cr
			  &\approx 4 \mathcal{A} \rho_{00} \mathcal{A}^\dagger / \kappa_b^2 + \mathcal{O}(\delta). 
\end{align}
Injecting the above expressions in Eq.~(\ref{B4}), we obtain 
\begin{align*}
	\dot{\rho}_a \approx \frac{4 |g|^2}{\kappa_b} \mathcal{D}[a^2] \rho_a + \frac{2 g \varepsilon_b}{\kappa_b} [a^{\dagger 2}-a^2,\rho_a] + \kappa_a \mathcal{D}[a]\rho_a. 
\end{align*}
Rearranging terms, we find 
\begin{align}\label{rho_a}
	\dot{\rho}_a &\approx \kappa_2 \mathcal{D}[a^2] \rho_a + \frac{\kappa_2 \alpha^2}{2} [a^{\dagger 2}-a^2,\rho_a] + \kappa_a \mathcal{D}[a]\rho_a \cr
				 &= \kappa_2 \mathcal{D}[a^2 - \alpha^2] \rho_a + \kappa_a \mathcal{D}[a]\rho_a, 
\end{align}
where $\kappa_2=4|g|^2/\kappa_b$ is the effective two-photon dissipation rate, $\alpha = \sqrt{\varepsilon_b / g}$ is the coherent amplitude. 

\begin{figure}
	\centering
	\scalebox{0.59}{\includegraphics{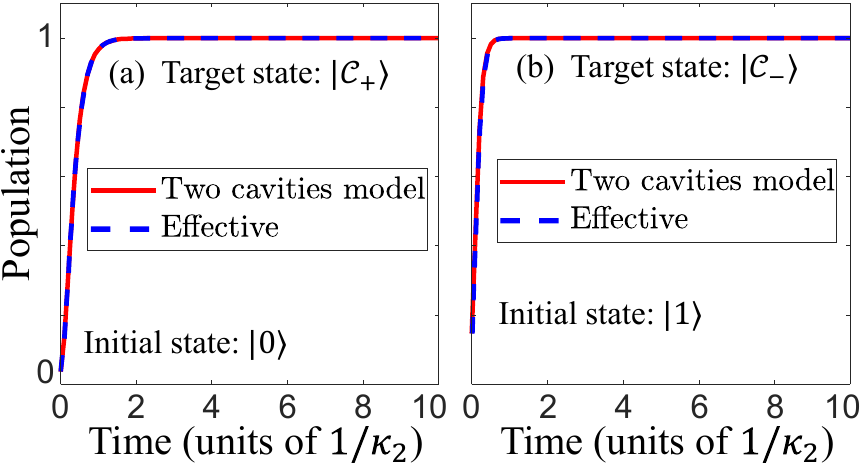}}
	\caption{Stabilization process of the system in the absence of damping in storage mode and pure dephasing. When the storage mode is in a low-excitation state, the 
			system spontaneously stabilizes into the cat state of the corresponding parity via two-photon dissipation. The red-solid curve denotes the dynamics evolution 
			governed by Eq.~(\ref{Rho}), and blue-dashed curve denotes that governed by Eq.~(\ref{rho_a}). (a) The system is initiated in vacuum state $|0\rangle$, the 
			target is the even cat-state $|\mathcal{C}_+\rangle$. (b) The system is initiated in Fock state $|1\rangle$, the target is the odd cat-state $|\mathcal{C}_-\rangle$. 
			We choose $\alpha = 2$, and this simulations are performed for $1 / \kappa_b = 25\ {\rm ns}$ and $g/2\pi = 111\ {\rm kHz}$ in Ref.~\cite{Leghtas2015}. }
	\label{TE}
\end{figure}

{As shown in Fig.~\ref{TE}, the steady state converges to the even cat-state $|\mathcal{C}_+\rangle$ when the system is initiated in the vacuum state (left panel), 
and to the odd cat-state $|\mathcal{C}_-\rangle$ when the system is initiated in the Fock state $|1\rangle$ (right panel). 
Figure \ref{TE} demonstrates that the effective description in Eq.~(\ref{rho_a}) accurately captures the system stabilization process described by Eq.~(\ref{Rho}). 
Therefore, the two-photon dissipation confinement can be represented by the effective dissipator $\kappa_2 \mathcal{D}[a^2 - \alpha^2] \rho$.}

\section{Optimal robustness condition derivation}\label{Apd_C}
The systematic error sensitivity in Sec.~\ref{Opt} reads 
\begin{align*}
	q_{s,\lambda} = \frac{1}{4} \left| \frac{m}{2} \int_{\theta_0}^{\theta_f} e^{-i m\theta} \sin2\theta d\theta + i\int_{\theta_0}^{\theta_f} e^{-i m\theta} d\theta \right|^2, 
\end{align*}
where $\theta_0 \equiv \theta(0) = 0$, $\theta_f \equiv \theta(t_f) = \pi$. 
In order to make $q_{s,\lambda}=0$, the following conditions need to be satisfied
\begin{align}
	&\int_{0}^{\pi} e^{-i m\theta} \sin2\theta d\theta = 0, \cr
	&\int_{0}^{\pi} e^{-i m\theta} d\theta = 0, 
\end{align}
For the first integral expression, 
\begin{align}
	&\int_{0}^{\pi} e^{-i m\theta} \sin2\theta d\theta \cr
	=& \frac{1}{2i} \int_{0}^{\pi} e^{-i m\theta} (e^{i 2\theta} - e^{-i 2\theta}) d\theta \cr
	=& \frac{1}{2i} \left[ \frac{1-(-1)^m}{i(m-2)} + \frac{(-1)^m-1}{i(m+2)} \right]. 
\end{align}
For the second expression, 
\begin{align}
	\int_{0}^{\pi} e^{-i m\theta} d\theta = \frac{1-(-1)^m}{im}. 
\end{align}
Therefore, when $m$ is even with $m \neq \pm2$, the systematic error sensitivity $q_{s,\lambda}=0$. 

{
\section{Derivation of the trade-off time window}\label{TW}
In the logical subspace, the effective Rabi frequency reads $\Omega_R(t) \approx 4 \alpha \varepsilon_d(t)$ for large $\alpha$. 
To ensure that the non-adiabatic drive does not overwhelm the dissipative confinement provided by the Liouvillian spectral gap $\Delta_L = 2\kappa_2 |\alpha|^2$, 
the drive amplitude must satisfy $\varepsilon_d(t) \ll \Delta_L$, which leads directly to the upper bound on the effective Rabi frequency 
\begin{align}
	\Omega_R(t) \ll 8\kappa_2 |\alpha|^3. 
\end{align}
To achieve complete population transfer from $|\mathcal{C}_+\rangle$ to $|\mathcal{C}_-\rangle$, the accumulated rotation angle on the Bloch sphere must satisfy 
\begin{align*}
	\int_0^{t_f} \Omega_R(t) dt \approx \pi. 
\end{align*}
Therefore, the time-averaged Rabi frequency must satisfy $\bar{\Omega}_R \approx \pi / t_f$. The average drive strength 
\begin{align}
	\bar{\varepsilon}_d \approx \frac{\bar{\Omega}_R}{4\alpha} \approx \frac{\pi}{4\alpha t_f} \sim \mathcal{O}\left(\frac{1}{\alpha t_f}\right). 
\end{align}
Imposing the confinement condition $\bar{\varepsilon}_d \ll \Delta_L$, we can obtain the lower bound on gate time, 
\begin{align}
	\frac{1}{\alpha t_f} \ll 2\kappa_2 |\alpha|^2 \implies t_f \gg \frac{1}{2\kappa_2 |\alpha|^3}. 
\end{align}

In the presence of single-photon loss $\kappa_a \mathcal{D}[a]\rho$, the projection onto the cat-state subspace $P_C a P_C \approx \alpha \sigma_x$ results in an 
effective bit-flip error channel $\kappa_a |\alpha|^2 \mathcal{D}[\sigma_x]\rho$ with rate $\Gamma_{\rm flip} \approx \kappa_a |\alpha|^2$. Over the duration $t_f$, 
the loss-induced infidelity scales as $\epsilon_{\rm flip} \approx \Gamma_{\rm flip} t_f \approx \kappa_a |\alpha|^2 t_f$. Requiring $\epsilon_{\rm flip} \ll 1$ 
imposes the upper bound 
\begin{align}
	t_f \ll \frac{1}{\kappa_a |\alpha|^2}.
\end{align}
Combining both limits defines the operational time window 
\begin{align}
	\frac{1}{2\kappa_2 |\alpha|^3} \ll t_f \ll \frac{1}{\kappa_a |\alpha|^2}. 
\end{align}
Therefore, operating within this trade-off time window is essential for achieving a high-fidelity population transfer, as it balances the suppression of non-adiabatic 
leakage against the accumulation of single-photon loss. 
}

{
\section{Decoherence sensitivity analysis}\label{qn}
To isolate the effect of decoherence, we assume that the systematic error in the control field is absent. We then analyze the sensitivity of the protocol to single-photon 
loss, which is one of the dominant logical error channels in cat-state qubits. The corresponding Lindblad master equation in the cat-state subspace reads 
\begin{align}\label{D1}
	\dot{\rho} = -i[H_c^{\rm eff},\rho] + \mathcal{D}[\gamma_s \sigma_x] \rho, 
\end{align}
where $\gamma_s = \sqrt{\kappa_a |\alpha|^2}$. The Bolch vector in Eq.~(\ref{r}) is 
\begin{align}\label{rr}
    \boldsymbol{r}(t) = \begin{pmatrix}
						   \sin\theta \cos\phi \cr
						   -\sin\theta \sin\phi \cr
                           \cos\theta
					    \end{pmatrix}. 
\end{align}
Substituting $\rho = \frac{1}{2} (1 + \boldsymbol{r} \cdot \boldsymbol{\sigma})$ and Eq.~(\ref{rr}) into Eq.~(\ref{D1}), we can obtain 
\begin{align}
    \dot{\boldsymbol{r}} = (L_c + L_s) \boldsymbol{r}, 
\end{align}
where 
\begin{align*}
    L_c = \begin{pmatrix}
            0 & -\Delta & 0 \cr
            \Delta & 0 & -\Omega_R \cr
            0 & \Omega_R & 0
          \end{pmatrix}, 
	\ \ 
    L_s = \begin{pmatrix}
            0 & 0 & 0 \cr
            0 & -2\gamma_s^2 & 0 \cr
            0 & 0 & -2\gamma_s^2
          \end{pmatrix}.
\end{align*}

The population of odd cat state $|\mathcal{C}_-\rangle$ at the time $t$ is $P_-(t) = \frac{1}{2} [1 - r_z(t)]$. Applying time-dependent perturbation theory, we can 
obtain \cite{Lu2013PRA}
\begin{align}
    r_z(t_f) \simeq -1 - \int_{0}^{t_f} dt \langle\boldsymbol{r}(t)|L_s|\boldsymbol{r}(t)\rangle, 
\end{align}
which results in 
\begin{align}
    P_-(t_f) \simeq& 1 - \gamma_s^2 \int_{0}^{t_f} (\sin^2\theta\sin^2\phi + \cos^2\theta) dt \cr
                =& 1 - \gamma_s^2 \int_{0}^{t_f} (1 - \sin^2\theta\cos^2\phi) dt. 
\end{align}
Defining the noise sensitivity as \cite{Lu2013PRA}
\begin{align}
    q_{n} = -\frac{1}{2} \frac{\partial^2 P_-}{\partial \gamma_s^2} \Bigg|_{\gamma_s=0}. 
\end{align}
Therefore, 
\begin{align}\label{q_n}
    q_{n} = \int_{0}^{t_f} (1 - \sin^2\theta\cos^2\phi) dt. 
\end{align}

It is worth noting that the noise sensitivity $q_n$ can be reduced through an appropriate choice of the control parameters $\theta(t)$ and $\phi(t)$, but it cannot be 
completely eliminated. Since 
\begin{align}
	0 \le \sin^2\theta\cos^2\phi \le 1, 
\end{align}
the integrand in Eq.~(\ref{q_n}) always satisfies
\begin{align}
	1 - \sin^2\theta\cos^2\phi \ge 0. 
\end{align}
The condition $q_n=0$ requires $\sin^2\theta\cos^2\phi = 1$ throughout the entire evolution, which implies $\theta = \pi/2$ and $\phi = 0$ (or $\pi$) for all time. 
However, the state transfer protocol requires the boundary conditions 
\begin{align*}
	\theta(0) = 0,\ \ \ \ \theta(t_f) = \pi. 
\end{align*}
Therefore, $q_n$ is always positive, indicating that the influence of single-photon loss cannot be completely eliminated. 
}

\bibliography{Ref}

\end{document}